\documentclass[lettersize,journal]{IEEEtran}
 
\usepackage[T1]{fontenc}
\usepackage{amsmath,amssymb}
\usepackage{graphicx}
\usepackage{cite}
\usepackage{url}
\usepackage{balance}
\usepackage{enumitem}
\setlist[enumerate]{leftmargin=*,nosep}

\usepackage{amsmath,amsfonts}
\usepackage{algorithm}
\usepackage{array}
\usepackage[caption=false,font=normalsize,labelfont=sf,textfont=sf]{subfig}
\usepackage{textcomp}
\usepackage{stfloats}       
\usepackage{verbatim}
\usepackage{graphicx}

\graphicspath{{figures/}}
 
\begin{document}
 
\title{Age of Information Under DCC Rate Constraints for V2I Broadcast Along Urban Corridors}
 
\author{Yousef~AlSaqabi
\thanks{Y.~AlSaqabi is with the Department of Electrical Engineering, Kuwait University, Kuwait City, Kuwait (e-mail: yousef.alsaqabi@ku.edu.kw).}%
}
 
\maketitle
 
% ---- Abstract ----
\begin{abstract}
ETSI Decentralized Congestion Control (DCC) limits roadside unit
(RSU) broadcast rates based on channel load, yet its impact on
age of information (AoI) for vehicle-to-infrastructure updates
remains uncharacterized under real traffic. We derive the AoI of
DCC-constrained V2I broadcast, revealing a hyperbolic density
dependence that induces diurnal AoI variation exceeding
$4\times$ on a four-RSU corridor, with the DCC target CBR
parameter as the dominant control. We propose a cooperative
policy exploiting upstream spatial traffic correlation to improve
channel load estimation, with a safeguard ensuring non-negative
gains. Evaluated on a 5-day, 762\,050-vehicle trace from Kuwait
City, the policy reduces corridor AoI by 5\% at moderate and up
to 66\% at conservative DCC settings.
\end{abstract}
 
% ---- Index Terms ----
 
% ======================================================================
% SECTION I: INTRODUCTION
% ======================================================================
\section{Introduction}
\label{sec:intro}
 Roadside units (RSUs) in intelligent transportation systems
broadcast status updates --- signal phase and timing, hazard
warnings, queue lengths --- to vehicles via ITS-G5 (IEEE
802.11p). The freshness of these updates, measured by the age of
information (AoI)~\cite{kaul2012infocom, yates2021jsac}, directly
affects the quality of cooperative awareness at receiving vehicles.
Current deployments use a fixed 10\,Hz broadcast rate regardless
of traffic conditions, resulting in an AoI of 100\,ms that is
insensitive to the actual channel state.

In ITS-G5 networks, ETSI Decentralized Congestion Control (DCC)
limits each station's transmit rate based on the measured channel
busy ratio (CBR)~\cite{etsi102687, etsi103175}. As vehicle density
rises, CBR increases and DCC reduces the allowed broadcast rate,
directly increasing AoI. Lyamin~\emph{et~al.}~\cite{lyamin2020networking}
showed that DCC rate restrictions degrade AoI at a single RSU.
Without DCC, the AoI-optimal strategy is trivially to transmit at
the maximum rate, since Maatouk~\emph{et~al.}~\cite{maatouk2020ton}
proved that the optimal CSMA backoff rate under symmetric channel
holding times is always the maximum allowed. DCC is therefore the
binding constraint that creates a nontrivial AoI optimization for
V2I broadcast. Despite this, no prior work has characterized the
AoI behavior of V2I broadcast under real DCC constraints with
measured time-varying traffic.

Existing AoI-aware scheduling work operates at a single RSU or base station with stationary traffic assumptions. Analytical AoI models for 802.11p broadcast have been derived at a single node~\cite{baiocchi2021cl, kaul2011secon}, convex CSMA backoff optimization has been studied for co-located links~\cite{maatouk2020ton}, scheduling policies with AoI guarantees have been derived for broadcast networks~\cite{kadota2018ton}, and optimal update policies have
been characterized for generate-at-will sources~\cite{sun2017tit}. DRL-based resource management has been proposed for single-cell V2X
systems~\cite{chen2020twc, abdelaziz2020tcom}. Multi-RSU collaboration has been considered for deadline-based scheduling~\cite{bok2017cluster} and unicast-multicast-broadcast
mode selection~\cite{alhabob2022globecom}, but neither work addresses AoI or DCC constraints. No prior work analyzes how AoI varies under DCC across a multi-RSU corridor with real diurnal traffic, nor proposes mechanisms to mitigate DCC-induced AoI degradation through inter-RSU cooperation.

This letter makes two contributions. First, we characterize the
AoI of DCC-constrained V2I broadcast along a multi-RSU urban
corridor. Using a 5-day traffic trace from four signalized
intersections in Kuwait City (762\,050 vehicles), we show that
DCC induces a diurnal AoI variation exceeding $4\times$ between
peak and trough hours, and that the DCC target CBR parameter
is the dominant factor controlling corridor AoI. Second, we
propose a cooperative mitigation policy in which each RSU
augments its local CBR measurement with an upstream traffic
observation, exploiting measured spatial correlations
($\rho > 0.85$) along the corridor. The policy includes a
safeguard ensuring it never degrades AoI below the reactive
baseline. Cooperative prediction reduces corridor AoI by 5\% at moderate DCC settings (up to 10\% at individual high-correlation RSUs), and by over 30\% under conservative configurations commonly used in safety-critical deployments.

% ======================================================================
% SECTION II: SYSTEM MODEL
% ======================================================================
\section{System Model and Problem Formulation}
\label{sec:model}

\subsection{Network Topology}

We consider a linear corridor of $N=4$ roadside units (RSUs) deployed at
signalized intersections along Second Ring Road, Kuwait City.
Intersections I1--I3 are four-way and I4 is three-way.
Each RSU broadcasts status updates (signal phase and timing,
hazard warnings, queue lengths) to vehicles within its coverage
area using ITS-G5 on a dedicated 10\,MHz channel in continuous
single-channel mode per ETSI EN~302~663~\cite{etsi302663}.
Adjacent RSUs are connected via wired backhaul, enabling
unidirectional upstream signaling: RSU~$i$ transmits its current
channel load observation to RSU~$i+1$.

\subsection{Traffic and Channel Load Model}

Vehicles arrive at RSU~$i$ according to a time-varying process
with hourly rate $\lambda_i(t)$ (veh/h). The number of vehicles
simultaneously present in the coverage area of RSU~$i$ follows
from Little's law:
\begin{equation}
  n_i(t) = \frac{\lambda_i(t)}{3600} \cdot \tau, \quad
  \tau = \frac{2R}{v},
  \label{eq:n_from_lambda}
\end{equation}
where $R=300$\,m is the coverage radius and $v=40$\,km/h is the
average vehicle speed, giving a sojourn time $\tau=54$\,s.

Each vehicle broadcasts cooperative awareness messages (CAMs) at
10\,Hz with frame duration $T_\mathrm{on}=498$\,$\mu$s
(300-byte payload at 6\,Mbps, including AIFS and PHY overhead)
\cite{etsi302637}. The channel busy ratio (CBR) at RSU~$i$,
measured over 100\,ms windows, is dominated by vehicle
transmissions:
\begin{equation}
  \mathrm{CBR}_i(t) = n_i(t) \cdot \frac{T_\mathrm{on}}{D_\mathrm{BSM}},
  \label{eq:cbr}
\end{equation}
where $D_\mathrm{BSM}=100$\,ms is the CAM generation period. Each
vehicle contributes approximately $T_\mathrm{on}/D_\mathrm{BSM}
\approx 0.5\%$ to the CBR. Table~\ref{tab:params} summarizes the
system parameters.

\begin{table}[t]
  \centering
  \caption{System Parameters}
  \label{tab:params}
  \footnotesize
  \begin{tabular}{lll}
    \hline
    \textbf{Parameter} & \textbf{Value} & \textbf{Source} \\
    \hline
    Slot time $\sigma$          & 13\,$\mu$s     & IEEE 802.11p \\
    AIFS (AC\_VO)               & 58\,$\mu$s     & IEEE 802.11p \\
    PHY preamble + SIGNAL       & 40\,$\mu$s     & IEEE 802.11p \\
    Data rate                   & 6\,Mbps        & ITS-G5 base rate \\
    CAM size                    & 300\,bytes     & ETSI EN 302 637-2 \\
    Frame duration $T_\mathrm{on}$ & 498\,$\mu$s & Computed \\
    CAM period $D_\mathrm{BSM}$   & 100\,ms     & ETSI EN 302 637-2 \\
    RSU coverage radius $R$     & 300\,m         & Typical urban \\
    Vehicle speed $v$           & 40\,km/h       & Urban arterial \\
    Sojourn time $\tau$         & 54\,s          & $2R/v$ \\
    \hline
    DCC target $\mathrm{CBR}_\mathrm{tgt}$ & 0.30--0.68 & ETSI TS 103 175 \\
    DCC threshold $\mathrm{CBR}_\mathrm{thr}$ & 0.10--0.15 & ETSI TS 102 687 \\
    DCC smoothing $\tau_\mathrm{DCC}$ & 1\,s & ETSI TS 102 687 \\
    \hline
    Number of RSUs $N$          & 4              & Deployment \\
    Spatial correlation $\rho$ & 0.848--0.977 & Measured \\
    \hline
  \end{tabular}
\end{table}

\subsection{ETSI DCC Rate Constraint}

ETSI Decentralized Congestion Control (DCC) limits each ITS
station's transmit rate based on the measured
CBR~\cite{etsi102687,etsi103175}. The maximum DCC-compliant
broadcast rate for the RSU is:
\begin{equation}
  \mu_\mathrm{max}(C) = \begin{cases}
    1/T_\mathrm{on}, & C \leq C_\mathrm{thr}, \\[4pt]
    (C_\mathrm{tgt} - C)/T_\mathrm{on}, &
      C_\mathrm{thr} < C < C_\mathrm{tgt}, \\[4pt]
    \mu_\mathrm{min}, & C \geq C_\mathrm{tgt},
  \end{cases}
  \label{eq:dcc}
\end{equation}
where $C \triangleq \mathrm{CBR}$, and $C_\mathrm{tgt}$ and
$C_\mathrm{thr}$ are the DCC target and threshold parameters.

The DCC constraint is the binding limitation on the RSU's
broadcast rate. Without DCC, the AoI-optimal strategy is trivially
to transmit at the maximum channel rate, since per-packet collision
probability remains below 2\% even at $n=120$ vehicles.
Maatouk~\emph{et~al.}~\cite{maatouk2020ton} proved that under
symmetric channel holding times, the AoI-optimal CSMA backoff rate
is always the maximum allowed (Lemma~1), confirming that an
external rate constraint is required to create a nontrivial
optimization.

\subsection{Age of Information}

The instantaneous AoI at a vehicle receiving updates from RSU~$i$ is
$\Delta_i(t) = t - U_i(t)$, where $U_i(t)$ is the generation
timestamp of the most recently received update~\cite{kaul2012infocom}.
The RSU operates in generate-at-will mode: it always has fresh
status data and transmits a newly generated update whenever it
gains channel access. Baiocchi~\emph{et~al.}~\cite{baiocchi2021itc}
showed that this buffer-less operation is AoI-optimal for 802.11p
broadcast. For a generate-at-will source broadcasting at effective
rate $\mu_i$ on a continuous channel, the average AoI
is~\cite{kaul2012infocom,yates2021jsac}
\begin{equation}
  \bar{\Delta}_i = \frac{1}{\mu_i}.
  \label{eq:aoi}
\end{equation}
Since $\bar{\Delta}_i$ is strictly decreasing in $\mu_i$, the
AoI-optimal rate is always the maximum DCC-compliant rate:
$\mu_i^*(t) = \mu_\mathrm{max}\bigl(\mathrm{CBR}_i(t)\bigr)$. The resulting sub-millisecond AoI at low traffic densities is well within the 100\,ms V2I latency budget specified by 3GPP~\cite{3gpp22185}.

\subsection{Corridor-Weighted AoI Objective}

The optimization objective is the time-averaged traffic-weighted
corridor AoI:
\begin{equation}
  \bar{\Delta}_\mathrm{corr} = \frac{1}{T} \int_0^T
    \frac{\sum_{i=1}^{N} \lambda_i(t)\,\bar{\Delta}_i(t)}
         {\sum_{i=1}^{N} \lambda_i(t)} \, dt,
  \label{eq:corridor_aoi}
\end{equation}
which weights each RSU's AoI contribution by its current traffic
share. This ensures that AoI reduction at high-traffic RSUs
contributes more to the objective than equivalent reduction at
low-traffic RSUs.

\subsection{Cooperative Upstream Signaling}

RSU~$i$ measures its current CBR and transmits this value to
RSU~$i+1$ via the wired backhaul. The propagation latency is
negligible relative to the timescale of traffic variation.
RSU~$i+1$ forms a predictive CBR estimate by combining its own
smoothed measurement with the upstream observation:
\begin{equation}
  \widehat{\mathrm{CBR}}_{i+1}(t) = \mathrm{CBR}_{i+1}^\mathrm{s}(t)
    + (1-\alpha)\,\rho_{i,i+1}\bigl[\mathrm{CBR}_i(t)
    - \mathrm{CBR}_{i+1}^\mathrm{s}(t)\bigr],
  \label{eq:coop_cbr}
\end{equation}
where $\mathrm{CBR}_{i+1}^\mathrm{s}(t)$ is the locally smoothed
CBR (exponential moving average with time constant
$\tau_\mathrm{DCC}$), $\rho_{i,i+1}$ is the spatial traffic
correlation between adjacent RSUs, and $\alpha \in (0,1)$ is a
mixing parameter. The cooperative rate is then
$\mu_{i+1}^\mathrm{coop}(t) = \mu_\mathrm{max}\bigl(
\widehat{\mathrm{CBR}}_{i+1}(t)\bigr)$.

The predictive estimate exploits the spatial correlation of
traffic flow along the corridor: vehicles passing RSU~$i$ will
arrive at RSU~$i+1$ after a propagation delay, so the upstream
CBR observation serves as a leading indicator of the downstream
channel load. When $\rho_{i,i+1}$ is high (ranging from 0.85 to 0.98 across adjacent pairs in our corridor), the upstream signal substantiallyimproves the CBR estimate relative to the local smoothed
measurement alone.

% ======================================================================
% SECTION III: COOPERATIVE DCC RATE OPTIMIZATION
% ======================================================================
\section{DCC-Constrained AoI Analysis and Cooperative Policy}
\label{sec:optimization}
 \subsection{AoI Under DCC Rate Constraints}

Substituting~\eqref{eq:cbr} into~\eqref{eq:dcc}
and~\eqref{eq:aoi}, the average AoI at RSU~$i$ operating at the
maximum DCC-compliant rate is
\begin{equation}
  \bar{\Delta}_i(t) = \begin{cases}
    T_\mathrm{on}, &
      n_i(t) \leq n_\mathrm{thr}, \\[4pt]
    \dfrac{T_\mathrm{on}}
      {\mathrm{CBR}_\mathrm{tgt} - n_i(t)\,T_\mathrm{on}/D_\mathrm{BSM}},
      & n_\mathrm{thr} < n_i(t) < n_\mathrm{tgt}, \\[6pt]
    1/\mu_\mathrm{min}, &
      n_i(t) \geq n_\mathrm{tgt},
  \end{cases}
  \label{eq:aoi_dcc}
\end{equation}
where $n_\mathrm{thr} = \mathrm{CBR}_\mathrm{thr}\,
D_\mathrm{BSM}/T_\mathrm{on}$ and
$n_\mathrm{tgt} = \mathrm{CBR}_\mathrm{tgt}\,
D_\mathrm{BSM}/T_\mathrm{on}$ are the vehicle counts at which
DCC activates and saturates, respectively.

Equation~\eqref{eq:aoi_dcc} reveals three operating regimes.
Below $n_\mathrm{thr}$, DCC is inactive and the RSU achieves its minimum AoI of Ton (i.e., $1/\mu_\mathrm{max} = T_\mathrm{on}$) $\approx 0.5$\,ms. In the active
DCC region, AoI increases nonlinearly with vehicle density due to
the hyperbolic dependence on $\mathrm{CBR}_\mathrm{tgt} -
\mathrm{CBR}_i$: as CBR approaches the target, AoI grows
sharply. Beyond $n_\mathrm{tgt}$, the RSU is throttled to a
minimum rate and AoI reaches its maximum. The sensitivity of AoI
to vehicle density is controlled by $\mathrm{CBR}_\mathrm{tgt}$:
a lower target tightens the active region and steepens the
AoI response.

\subsection{Reactive DCC Baseline}

Under standard DCC operation, each RSU independently measures its
local CBR and applies exponential smoothing:
\begin{equation}
  \mathrm{CBR}_i^\mathrm{s}(t) = (1 - \alpha_\mathrm{s})\,
    \mathrm{CBR}_i^\mathrm{s}(t - \Delta t)
    + \alpha_\mathrm{s}\,\mathrm{CBR}_i(t),
  \label{eq:cbr_smooth}
\end{equation}
where $\alpha_\mathrm{s} = \Delta t / \tau_\mathrm{DCC}$ is the
smoothing coefficient. The reactive policy sets the broadcast rate
to the DCC-compliant maximum based on this smoothed measurement:
\begin{equation}
  \mu_i^\mathrm{react}(t) = \mu_\mathrm{max}\bigl(
    \mathrm{CBR}_i^\mathrm{s}(t)\bigr).
  \label{eq:mu_react}
\end{equation}
This policy is AoI-optimal given the local smoothed CBR estimate;
any deviation from the true CBR causes a rate mismatch that
increases AoI relative to the oracle policy
$\mu_i^* = \mu_\mathrm{max}(\mathrm{CBR}_i(t))$.

\subsection{Cooperative Predictive DCC}

The cooperative policy augments the local smoothed CBR with the
upstream RSU's current observation. From~\eqref{eq:coop_cbr},
RSU~$i{+}1$ computes the predictive estimate
$\widehat{\mathrm{CBR}}_{i+1}(t)$ and sets its rate as
\begin{equation}
  \mu_{i+1}^\mathrm{coop}(t) = \max\!\Big(
    \mu_\mathrm{max}\bigl(\widehat{\mathrm{CBR}}_{i+1}(t)\bigr),\;
    \mu_{i+1}^\mathrm{react}(t)\Big).
  \label{eq:mu_coop}
\end{equation}
The $\max$ operator ensures the cooperative policy never reduces
the broadcast rate below the reactive baseline. If the upstream
signal produces a less favorable estimate than the local
measurement, the RSU defaults to the reactive rate. For the head
RSU ($i=1$), no upstream signal is available and the policy
reduces to reactive DCC.

\subsection{Cooperation Gain Analysis}

The instantaneous AoI reduction at RSU~$i$ from cooperation is
\begin{equation}
  G_i(t) = \frac{1}{\mu_i^\mathrm{react}(t)}
          - \frac{1}{\mu_i^\mathrm{coop}(t)} \geq 0.
  \label{eq:gain_inst}
\end{equation}
In the active DCC region, substituting the linear $\mu_\mathrm{max}$
from~\eqref{eq:dcc} yields
\begin{equation}
  G_i(t) = T_\mathrm{on}\!\left(
    \frac{1}{\mathrm{CBR}_\mathrm{tgt} - \mathrm{CBR}_i^\mathrm{s}}
  - \frac{1}{\mathrm{CBR}_\mathrm{tgt} - \widehat{\mathrm{CBR}}_i}
  \right)\!,
  \label{eq:gain_expanded}
\end{equation}
where time arguments are suppressed for clarity. Three factors
govern the magnitude of this gain.

\emph{1) DCC conservativeness.}
From~\eqref{eq:aoi_dcc}, the AoI sensitivity to CBR estimation
errors increases as $\mathrm{CBR}_\mathrm{tgt}$ decreases,
because the hyperbolic $1/(\mathrm{CBR}_\mathrm{tgt} -
\mathrm{CBR}_i)$ term steepens near the DCC saturation point.
Conservative DCC settings, common in safety-critical deployments,
therefore amplify the benefit of improved CBR estimation.

\emph{2) Spatial correlation.}
The cooperative correction in~\eqref{eq:coop_cbr} is proportional
to $\rho_{i-1,i}$. Higher spatial correlation between adjacent
RSUs produces a larger and more accurate correction, while
$\rho = 0$ reduces the cooperative policy to the reactive
baseline.

\emph{3) Traffic dynamics.}
The gain concentrates during traffic transition periods when the
locally smoothed CBR deviates from the true channel state. During
steady traffic, the smoothed measurement converges and the
upstream signal provides diminishing additional information.

The corridor-level cooperation gain is
\begin{equation}
  G_\mathrm{corr} = \frac{\bar{\Delta}_\mathrm{corr}^\mathrm{react}
    - \bar{\Delta}_\mathrm{corr}^\mathrm{coop}}
    {\bar{\Delta}_\mathrm{corr}^\mathrm{react}},
  \label{eq:gain_corridor}
\end{equation}
where $\bar{\Delta}_\mathrm{corr}^\mathrm{react}$ and
$\bar{\Delta}_\mathrm{corr}^\mathrm{coop}$ are defined
per~\eqref{eq:corridor_aoi}. Since $G_1(t)=0$, the corridor
gain is driven by the downstream RSUs weighted by their
traffic share.

% ======================================================================
% SECTION IV: EVALUATION
% ======================================================================
\section{Evaluation}
\label{sec:evaluation}
\subsection{Setup}

The model is evaluated on a 5-day traffic trace (August 15--19,
2025) from four signalized intersections on Second Ring Road,
Kuwait City. Intersections I1--I3 are four-way and I4 is
three-way. Weekend days follow the Gulf convention
(Friday--Saturday). Table~\ref{tab:traffic} summarizes the
traffic characteristics. Spatial correlations between adjacent
RSUs, computed from the 120 hourly observations (5 days $\times$
24 hours), range from $\rho_{3,4}=0.848$ to $\rho_{1,2}=0.977$.

\begin{table}[t]
  \centering
  \caption{Traffic Trace Summary (Second Ring Road, Kuwait City)}
  \label{tab:traffic}
  \footnotesize
  \begin{tabular}{lcccc}
    \hline
    & \textbf{I1} & \textbf{I2} & \textbf{I3} & \textbf{I4} \\
    \hline
    Weekday daily (veh)   & 41\,905 & 50\,205 & 42\,149 & 42\,072 \\
    Weekend daily (veh)   & 27\,828 & 36\,630 & 31\,963 & 20\,108 \\
    Peak hour (weekday)   & 20:00   & 14:00   & 19:00   & 18:00   \\
    Peak/trough ratio     & 15.2$\times$ & 13.7$\times$
                          & 14.2$\times$ & 24.6$\times$ \\
    Upstream $\rho$       & ---     & 0.977   & 0.947   & 0.848   \\
    \hline
  \end{tabular}
\end{table}

Hourly vehicle counts from each intersection are summed across
all approach directions and interpolated to one-second resolution.
Since DCC operates at 100\,ms measurement windows, sub-hourly
traffic fluctuations from signal cycles and platoon arrivals
affect the instantaneous CBR. These are modeled as an
Ornstein--Uhlenbeck process (correlation time 300\,s, coefficient
of variation 0.10) superimposed on the interpolated hourly
profile. Each of the five days is simulated individually; reported
averages are computed across per-day results. The DCC smoothed CBR
is updated each second via~\eqref{eq:cbr_smooth} with
$\tau_\mathrm{DCC}=1$\,s, and the mixing parameter is
$\alpha=0.5$. The total trace spans 762\,050 vehicles across
the corridor.

Three policies are compared:
\begin{enumerate}
  \item \textbf{Fixed 10\,Hz}: the industry-standard periodic
    rate, subject to DCC rate limiting.
  \item \textbf{Reactive DCC}: each RSU independently sets
    $\mu_i^\mathrm{react}(t) = \mu_\mathrm{max}(
    \mathrm{CBR}_i^\mathrm{s}(t))$ per~\eqref{eq:mu_react}.
  \item \textbf{Cooperative DCC}: each downstream RSU uses
    the upstream signal per~\eqref{eq:mu_coop}.
\end{enumerate}

\subsection{DCC-Induced AoI Variation}

\begin{figure}[t]
  \centering
  \includegraphics[width=\columnwidth]{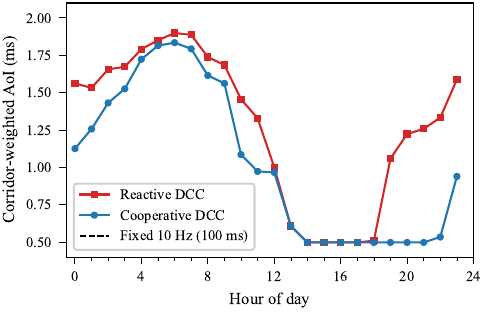}
  \caption{Corridor-weighted AoI over a representative weekday
    ($\mathrm{CBR}_\mathrm{tgt}=0.50$,
    $\mathrm{CBR}_\mathrm{thr}=0.15$). DCC induces a diurnal AoI
    variation exceeding $4\times$ between trough and peak hours.
    The cooperative policy reduces AoI during transition periods.
    Fixed 10\,Hz yields 100\,ms (off-scale).}
  \label{fig:hourly_aoi}
\end{figure}

Fig.~\ref{fig:hourly_aoi} shows the corridor-weighted AoI over
a representative weekday (Sunday) for $\mathrm{CBR}_\mathrm{tgt}
=0.50$ and $\mathrm{CBR}_\mathrm{thr}=0.15$. The fixed 10\,Hz
policy yields a constant AoI of 100\,ms (off-scale), confirming
that traffic-agnostic broadcasting wastes available channel
capacity.

Both adaptive policies reduce AoI by two orders of magnitude,
but the diurnal traffic pattern induces a pronounced AoI
variation: corridor AoI ranges from approximately 0.50\,ms during
low-traffic hours (when DCC is inactive and the RSU transmits at
the full channel rate of $1/T_\mathrm{on} \approx 2008$\,Hz) to
nearly 2.0\,ms during peak hours (when DCC restricts the rate),
a variation exceeding $4\times$. This variation is a direct
consequence of the nonlinear AoI--density relationship
in~\eqref{eq:aoi_dcc}: as vehicle density pushes CBR into the
active DCC region, AoI increases hyperbolically.

The cooperative policy tracks below reactive DCC throughout the
day. The gap is most visible during traffic transition hours and
vanishes during the trough when DCC is inactive at both policies.
Averaged over the five-day trace, the corridor AoI is 0.998\,ms
under reactive DCC and 0.946\,ms under cooperative DCC, a
reduction of 5.2\%.

\subsection{Impact of DCC Conservativeness}

\begin{figure}[t]
  \centering
  \includegraphics[width=\columnwidth]{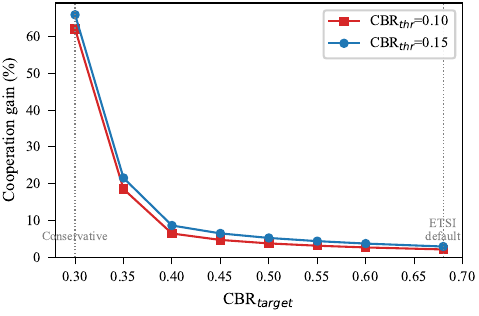}
  \caption{Cooperation gain versus DCC target CBR, averaged over
    the 5-day trace. Conservative DCC settings yield the largest
    gains due to the heightened AoI sensitivity to CBR
    estimation errors near the DCC saturation point.}
  \label{fig:cbr_sweep}
\end{figure}

Fig.~\ref{fig:cbr_sweep} shows the cooperation gain as a
function of $\mathrm{CBR}_\mathrm{tgt}$ for two values of
$\mathrm{CBR}_\mathrm{thr}$, averaged across all five days.
The gain increases monotonically as $\mathrm{CBR}_\mathrm{tgt}$
decreases, confirming the analytical prediction
from~\eqref{eq:gain_expanded}: a tighter DCC constraint amplifies
the AoI sensitivity to CBR estimation errors. At the permissive
ETSI default ($\mathrm{CBR}_\mathrm{tgt}=0.68$), the cooperative
gain is 2.9\%. Under conservative settings
($\mathrm{CBR}_\mathrm{tgt}=0.30$), the gain reaches 65.9\% because at this setting the DCC constraint approaches saturation during peak hours, and even modest CBR improvements prevent the RSU from being throttled to its minimum rate.

This result has practical implications: safety-critical V2I
deployments, which typically adopt conservative DCC
settings~\cite{lyamin2020networking} to maintain low channel
occupancy, stand to benefit most from cooperative upstream
signaling.

\subsection{Effect of Spatial Correlation and Per-RSU Breakdown}

\begin{figure}[t]
  \centering
  \includegraphics[width=\columnwidth]{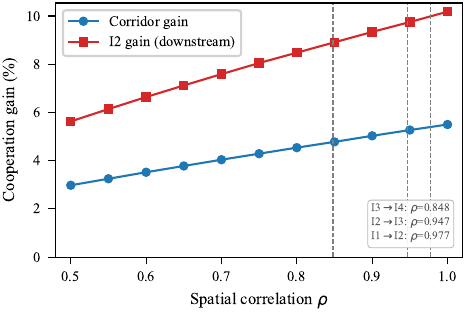}
  \caption{Cooperation gain versus spatial correlation~$\rho$
    ($\mathrm{CBR}_\mathrm{tgt}=0.50$). Dashed lines mark the
    measured values for the three adjacent RSU pairs.}
  \label{fig:rho_sweep}
\end{figure}

Fig.~\ref{fig:rho_sweep} shows the cooperation gain as a
function of~$\rho$ at $\mathrm{CBR}_\mathrm{tgt}=0.50$. Both the
per-RSU gain at~I2 and the corridor gain increase monotonically
with~$\rho$. The three measured correlation values are marked on
the figure.

Table~\ref{tab:per_rsu} presents the per-RSU breakdown averaged
over the five-day trace. I2 ($\rho=0.977$) achieves the highest
gain at 9.6\%, while I4 ($\rho=0.848$) achieves 5.0\%. RSU~I1,
as the corridor head, receives no upstream signal and has zero
cooperation gain by construction. I3 shows a small gain (1.4\%)
despite its moderate correlation with I2, because its peak
traffic hours differ from I2's (19:00 vs.\ 14:00), limiting the
predictive value of the upstream signal during I3's own transition
periods. The per-day ranges in Table~\ref{tab:per_rsu} confirm
that the gains are consistent across all five days of the trace.

\begin{table}[t]
  \centering
  \caption{Per-RSU Cooperation Gain
    ($\mathrm{CBR}_\mathrm{tgt}=0.50$, 5-Day Average)}
  \label{tab:per_rsu}
  \footnotesize
  \begin{tabular}{lccc}
    \hline
    \textbf{RSU} & \textbf{Upstream $\rho$} &
    \textbf{Gain (\%)} &
    \textbf{Per-day range (\%)} \\
    \hline
    I1 (head)  & ---   &  0.0 & --- \\
    I2         & 0.977 &  9.6 & 7.1--13.8 \\
    I3         & 0.947 &  1.4 & 0.0--1.8 \\
    I4         & 0.848 &  5.0 & 0.0--6.2 \\
    \hline
    Corridor   & ---   &  5.2 & 3.7--6.2 \\
    \hline
  \end{tabular}
\end{table}

The cooperation gain is insensitive to the DCC smoothing time
constant $\tau_\mathrm{DCC}$ across the range 0.5--5.0\,s,
varying by less than 0.5 percentage points. This confirms that
the gain derives from spatial prediction of the slowly varying
diurnal traffic component rather than from overcoming per-second
smoothing lag. The sub-hourly noise coefficient of variation has a
moderate effect: reducing it to zero (pure hourly interpolation)
yields a corridor gain of 4.3\%, while increasing it to 0.20
raises the gain to 7.6\%, bounding the sensitivity to this
modeling assumption.
 
% ======================================================================
% SECTION V: CONCLUSION
% ======================================================================
\section{Conclusion}
\label{sec:conclusion}
This letter characterized the age of information of
DCC-constrained V2I broadcast along a multi-RSU urban corridor.
Using a 5-day traffic trace from four intersections in Kuwait
City, we showed that ETSI DCC induces a diurnal AoI variation
exceeding $4\times$ between peak and trough hours, with the DCC
target CBR parameter acting as the dominant control on corridor
AoI. We proposed a cooperative policy in which each RSU augments
its local CBR measurement with an upstream traffic observation to
set its DCC-compliant broadcast rate. The policy reduces corridor
AoI by 5.2\% at moderate DCC settings
($\mathrm{CBR}_\mathrm{tgt}=0.50$) and by up to 65.9\% under
conservative configurations, with per-RSU gains scaling with the
measured spatial traffic correlation. The cooperative gain is
robust to the DCC smoothing time constant, confirming that the
benefit derives from spatial prediction of the diurnal traffic
component. Future work includes extension to adaptive DCC
(LIMERIC), longer multi-hop corridors, and C-V2X sidelink
environments.
 
% ======================================================================
% REFERENCES
% ======================================================================

\bibliographystyle{IEEEtran}
\bibliography{references}

\end{document}